\documentclass[10pt,conference]{IEEEtran}
\IEEEoverridecommandlockouts
\usepackage{cite}
\usepackage{amsmath,amssymb,amsfonts}
\usepackage{graphicx}
\usepackage{textcomp}
\usepackage{xcolor}
\def\BibTeX{{\rm B\kern-.05em{\sc i\kern-.025em b}\kern-.08em
    T\kern-.1667em\lower.7ex\hbox{E}\kern-.125emX}}

\usepackage{savesym}

\usepackage{graphicx}
\usepackage{float}
\usepackage{epstopdf}
\usepackage{subcaption}
\usepackage{booktabs} 
\usepackage{listings}
\usepackage{xcolor}
\usepackage{xcolor}

\usepackage{multicol}
\usepackage{multirow}
\definecolor{CPPLight}  {HTML} {686868}
\definecolor{CPPSteel}  {HTML} {888888}
\definecolor{CPPDark}   {HTML} {262626}
\definecolor{CPPBlue}   {HTML} {4172A3}
\definecolor{CPPGreen}  {HTML} {487818}
\definecolor{CPPBrown}  {HTML} {A07040}
\definecolor{CPPRed}    {HTML} {AD4D3A}
\definecolor{CPPViolet} {HTML} {7040A0}
\definecolor{CPPGray}  {HTML} {B8B8B8}

\makeatletter
\newif\if@restonecol
\makeatother

\usepackage[linesnumbered,ruled,vlined]{algorithm2e}
\usepackage{algpseudocode}

\usepackage{authblk}

\author[1]{Chen Tian}
\author[1]{Yazhe Wang}
\author[2]{Peng Liu}
\author[1]{Ruirui Dai}
\author[1]{Anyuan Zhou}
\author[1]{Xinwang Zhuo}
\affil[1]{State Key Laboratory of Information Security, Institute of Information Engineering, Chinese Academy of Sciences}
\affil[2]{College of Information Sciences and Technology, Pennsylvania State University, University Park}

\begin{document}

\title{Prihook: Differentiated context-aware hook placement for different owners' smartphones}

\maketitle

\begin{abstract}
A hook is a piece of code. It checks user privacy policy before some sensitive operations happen. We propose an automated solution named Prihook
for hook placement in the Android Framework. Addressing specific context-aware user privacy concerns, the hook placement in {\sf Prihook} is personalized. Specifically, we design User Privacy Preference Table (UPPT) to help a user express his privacy concerns. And we leverage machine learning to discover a Potential Method Set (consisting of Sensor Data Access Methods and Sensor Control Methods) from which we can select a particular subset to
put hooks. We propose a mapping from words in the UPPT lexicon to methods in the
Potential Method Set. With this mapping, {\sf Prihook} is able to (a) select a specific set of methods; and (b) generate and place hooks automatically. We test {\sf Prihook} separately on 6 typical UPPTs representing 6 kinds of resource-sensitive UPPTs, and no user privacy violation is found. The experimental results show that the hooks placed by {\sf PriHook} have small runtime overhead.
\end{abstract}

\section{Introduction}\label{intro}
A typical smartphone usually employs a spectrum of sensors. These sensors are able to provide information about location, sounds, images in a user surrounding environment. At the OS level, these sensors are shared resources that can be used by different applications. User apps are often able to access sensor resources through system services (e.g., \texttt{SensorService}, \texttt{CameraService}). For example, a car-hailing app (e.g., Uber) can call \texttt{LocationManagerService} to obtain a user's GPS and then use it to search the available taxies nearby.

Several studies have already shown that if the above resources are involved in a user privacy concern, a user app may leak sensitive personal data through abusing involved sensor resource \cite{owusu2012accessory,liu2015exploring,michalevsky2014gyrophone,narain2016inferring,templeman2014placeavoider,mohamed2016smashed,liu2015good}. A privacy concern can be viewed as a combination of a resource, a particular context in which the resource is accessed, and a policy about how to deal with the accesses. A context can be represented by a set of information items like time, date, location, system status. For example, when a user is in a private meeting context (e.g., a private business negotiation in Hotel X at 9 a.m.), he is always concerned about his conversation and
physical location and may expect to disable GPS, camera and microphone. So in this case, his privacy concern involves the particular resources (i.e., GPS, microphone and camera), the particular policy (i.e., disable GPS, camera and microphone) and the particular context (i.e., the system time of smartphone is 9 a.m. and the GPS position is Hotel X). Further, different users may have different privacy concerns and adopt different privacy policies. Accordingly, no universally applicable policy really exists to meet the privacy protection needs of all users.

To address the individual privacy concerns, on Android smartphones, the existing permission mechanism allows a user to accept or deny permissions on sensitive resources (e.g., camera, microphone) at app installation time or at runtime. However, researchers still try to place hooks into the Android Framework (i.e., system services) for privacy protection due to the following reasons: First, the existing Android permission mechanism provides an unlimited use of user-approved permissions for apps once installed. When
a resource is accessed, there is no context-aware
check in the standard Android Framework. However, prior studies \cite{Ghosh2011Privacy,Felt2012Android} have shown that users often change their perceptions about what permissions should be approved to an app when apprised of the various sensitive contexts. Hence, comparing the existing one-time permission granting, it requires access restrictions based on user privacy concerns at runtime.
Second, resources
are directly accessed by methods in the Framework.
If we want to conduct context-aware checks, placing
hooks before these access methods is a straightforward
way.

Listing \ref{list:code} shows a method with a hook in ipShield \cite{chakraborty2014ipshield}. This method is sensitive because it may send GPS data to a user app. ipShield adds a hook in it to protect GPS data. When a user app calls this method, the hook will check the context-aware privacy policy for the app and decide whether this method can or cannot access the current GPS resource (Line 7). If the access is disallowed, it will ignore the request and start checking the access request of the next app in the queue (Line 11). Otherwise, it does nothing but prints a log entry (Line 8). Before data is actually sent to a user app, ipShield checks the policy again to control the data granularity (e.g., how big the location radius should be) using a designated form of data transformation (Line 13). Therefore, a hook is a piece of code. It checks user privacy
policy before some sensitive operations happen.

\vspace*{-2.5mm}
\begin{lstlisting}[
caption={A method with a hook in the real world},
label={list:code},
]
private void handleLocationChangedLocked(Location location, boolean passive) {
    //some existing code in handleLocationChangedLocked
//The code from line 4 to line 14 is a hook.
for (UpdateRecord r : records) {
    RuleKey ruleKey = new RuleKey(TYPE_GPS, receiver.mUid, receiver.mPackageName);
    Rule rule = mPrivacyRules.get(ruleKey);
    if (mSensorPerturb.isActionPlayback(rule)) {
	Log.d(TAG, "Sending Playback location to " + receiver.mPackageName);
	} else {
	Log.d(TAG, "Not sending Playback location to " + receiver.mPackageName);
	continue;
	}
notifyLocation = mSensorPerturb.transformData(notifyLocation, rule);
}//end of a hook
    //The rest of this method sends GPS data to a user app.
}
\end{lstlisting}

Some existing works \cite{olejnik2017smarper,conti2010crepe} start from {\em permission-checks} to place hooks. However, our key observation is that although the code in the Android Framework can check whether a user app has permission to use some resources, it often checks permission at the beginning of requesting a resource (such as new an object or a thread). Once it grants that permission to an app, there will be no check during use of that resource. If we only add hooks in permission-checks, which means no context-aware check exists during use, privacy violation may happen. For instance, permission-checks can decide whether an app has permission to record audio through microphone at the beginning of opening a record. However, during recording, there is no permission check in the existing Android Framework. Hence, no context-aware check will be enforced because we only add hooks in permission-checks. Once the context has changed after recording for a while, privacy violation can happen because the changed context may not allow recording any more. Hence, placing hooks in permission-checks is not a good choice for privacy protection. Meanwhile, onboard sensors are {\em zero-permissioned}. Methods related to them do not involve any permission check. Someone who uses permission-checks as a clue for hook placement will omit these methods.

Although several efforts have attempted to place hooks in the Android Framework, these efforts
have been beset with some typical mistakes. Table \ref{tab:hookmistake} shows some hook mistakes in the existing context-aware solutions. The definition of mistakes will be discussed shortly in Section \ref{motiv}.

\textbf{Problem Statement.} \emph{How to automate the placement of context-aware hooks in the Android Framework based on specific context-aware user privacy concerns and
avoid the three mistakes (i.e., the bypass, the no-isolation,
the useless)?}

The main goal of this paper is to provide an automated, personalized and scalable solution named {\sf Prihook} for hook placement in the Android Framework. Because an expert cannot ensure that the attacker will never be able to find a creative way to bypass the deployed hooks and violate the privacy policy, our solution does not guarantee it either. We implement and evaluate {\sf Prihook} based on 3 key ideas. First, before placing hooks, we should systematically inspect the Android Framework code and discover a \emph{potential} set of methods from which we can select a particular subset to put hooks \textbf{(K1)}. One main contribution of this work is that we search and analyze the whole Android Framework to provide the Potential Method Set (PMS). This set consists of two kinds of methods: Sensor Data Access Method (SDAM) and Sensor Control Method (SCM) (see Section \ref{concept}). Second, we need to design user-friendly interface to help a user express privacy concerns clearly \textbf{(K2)}. Third, the hook placement should be automated and personalized based on specific context-aware user privacy concerns \textbf{(K3)}.

\textbf{Challenges.} To implement these three ideas, we are facing three main challenges. First, considering the large code base of the Android Framework, manual inspection of the code is often impractical. How can we automatically and scalably discover a potential set of methods that are appropriate for the hook placement (\textbf{C1})? One may want to use taint tracking or static analysis techniques to discover all privacy-relevant methods. However, we found that existing taint analysis or static analysis techniques/tools fall short of addressing this challenge. In Section \ref{ml}, we will give a new insight on why taint tracking or static analysis techniques fail to discover SDAMs and SCMs. Second, most mobile users lack professional knowledge about security and each one's privacy concerns may be different. How can we design an interface which is simple and efficient to describe hundreds of millions of users' context-aware privacy concerns (\textbf{C2})? Third, how can we design an efficient approach to identify the specific set of SDAMs and SCMs when a specific user privacy concern (including
resources, context and policies) is defined (\textbf{C3})?

To address Challenge C1, we leverage machine learning (i.e., \emph{support vector machines} in our case) to discover the Potential Method Set. To address Challenge C2, based on two principles (i.e., good coverage and easy to use), we design a table named User Privacy Preference Table (UPPT) and implement it in a portal app. The app helps a user express his or her privacy concerns by choosing a bunch of simple words from a portal UI. Then, based on the user privacy decisions, a UPPT will be generated for hook placement. To address Challenge C3, one key observation is that, given a specific user privacy concern, we often cannot directly select methods relevant to privacy concerns from the Potential Method Set.
Hence, we build a mapping from words in a UPPT to methods in the Potential Method Set. With this mapping, {\sf Prihook} is able to (a) select a specific set of methods; and (b) generate and place hooks automatically and personally.

\begin{table}[]
\centering
\fontsize{8}{10}\selectfont
\caption{Hook mistakes in the existing context-aware privacy protections}
\label{tab:hookmistake}
\begin{tabular}{|c|c|c|}
\hline
\textbf{Context-aware privacy protections}     & \textbf{Hook mistakes} \\ \hline
ipShield\cite{chakraborty2014ipshield} & bypass                                          \\ \hline
Semadroid\cite{xu2015semadroid}              & no-isolation                                          \\ \hline
Viola\cite{Mirzamohammadi2016Viola}\&Aurasium\cite{Xu2012Aurasium}\&AppOp\cite{appops}          & useless                                          \\ \hline
Protect My Privacy (PMP)\cite{PMP}             & bypass, no-isolation                                          \\ \hline
\end{tabular}
\vspace*{-4mm}
\end{table}

In summary, our main contributions are as follows:
\begin{itemize}
\item
    We design UPPT which can help a user define his own privacy concerns (including resources, context, and policies) easily and efficiently. We introduce a semantic intermediate layer named Operation Abstract Layer (OAL) and propose a mapping from words in the UPPT lexicon to methods in the Potential Method Set (PMS).
\item
    We implement a hooking tool named {\sf PriHook} which can automate the hook placement in the Android Framework. Based on a specific UPPT, {\sf Prihook} can select methods from the PMS, generates and puts hooks for the selected methods automatically.
\item
    We perform a thorough evaluation of {\sf PriHook}. We test it separately on 6 typical UPPTs and no user privacy violation is found. The experimental results show that the hooks placed by {\sf PriHook} have small runtime overhead.
\end{itemize}

\section{Background}

\begin{figure}
\centering
\includegraphics[width=0.48\textwidth,height=4.9cm]{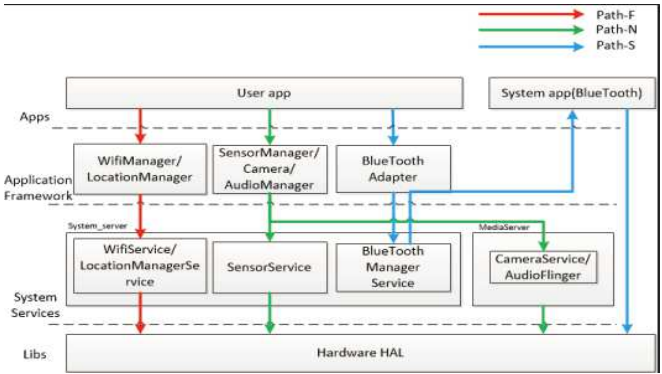}
\caption{The sensor dataflow on Android smartphones}
\label{figure:path}
\vspace*{-4.5mm}
\end{figure}

\subsection{Existing Sensor Dataflow on Android Smartphones}
Different sensors have different data paths. We group six classes of sensors into 3 typical data paths as shown in Figure \ref{figure:path}. Path-F is used by wifi, GPS. Path-N is used by onboard sensors (e.g., accelerometer and gyroscope), camera and microphone. Path-S is used by bluetooth.

Typically, all the services in Figure \ref{figure:path} are system services, each of which runs as a separate thread within the \texttt{system}\_\texttt{server} process and starts at system boot time. The difference between them is focused on whether it is a Java thread or a C++ thread. For instance, \texttt{LocationManagerService}, and \texttt{WifiService} are implemented in Android Java Framework and work as Java threads. \texttt{SensorService} is implemented as a native service and works as a C++ thread. \texttt{CameraService} and \texttt{AudioFlinger} are special cases. They run as native C++ threads in the \texttt{MediaServer} process.

Apps typically do not communicate with the services directly. Each service has a corresponding proxy named \texttt{Manager} such as \texttt{LocationManager} and \texttt{SensorManager}. Hence, apps can access sensor data through calling a \texttt{Manager} object's public methods. Both apps and the \texttt{Manager} object run in the same process of a Dalvik (or ART) Virtual Machine. Onboard sensors, GPS, microphone and wifi are in this case. Camera and bluetooth are special cases. For instance, a user app can use a \texttt{BlueToothAdapter} object to request operations such as discovering or enabling bluetooth.

\subsection{Android Access Control Mechanism}
In Android, sensitive resources such as sensor data are protected by a permission-based access control system. Unlike Linux, Android system does not employ a single checkpoint as reference monitor to check all the permissions. The permission checks spread through the Android Framework code and may be called when a SDAM or a SCM is invoked. However, not all access of resources results in a permission check. For instance, onboard sensors (e.g., proximity sensor, accelerometer) are {\em zero-permission}, which means no permission check exists in corresponding SDAMs and SCMs.

Google develops SEAndroid by porting SElinux's type enforcement (TE) MAC policy to the Android platform \cite{smalley2013security}. It enforces mandatory policy on system-level operations between subjects and objects (e.g., system calls). Processes are often subjects and files are regarded as objects. Besides, subjects and objects are labeled with a security context. The subject label is called a \emph{domain}, and the object label is called a \emph{type}. SEAndroid policies define which domain of subjects can access which type of objects with a bunch of permissions \cite{Loscocco2001Integrating}.

\section{Motivation}\label{motiv}
Investigating many solutions \cite{xu2015semadroid,chakraborty2014ipshield,Mirzamohammadi2016Viola,roesner2014world,Xu2012Aurasium,Petracca2015AuDroid,XposedFramework,PMP,appops}, we find that three kinds of mistakes may happen when a developer uses his expertise to put hooks. A key motivation of this work is that we intend to provide an automated hooking tool to reduce the chance of making mistakes.

\emph{Class A (bypass)}: Hooks are bypassed by a user app. For instance, ipShield puts a hook in function \texttt{handleLocationChangedLocked} to check policy for a user request of GPS data. However, it omits \texttt{getLastLocation}. This method can also be used by a user app to obtain the last location information.

\emph{Class B (no-isolation)}: A hook executes in the memory of a user app process. For instance, Semadroid \cite{xu2015semadroid} puts hooks in \texttt{ListnerDelegate} to control the accuracy of onboard sensors' data that a user app can obtain. However, the \texttt{ListnerDelegate} is implemented as a component in a user app. Unlike a hook which runs in a process in system services or apps, a no-isolation hook is in user apps' process and hence are prone to be modified by user apps.

\emph{Class C (useless)}: Placing hooks in the location where no privacy-relevant resource needs to be protected. This is a common mistake in security solutions \cite{Mirzamohammadi2016Viola,roesner2014world,Xu2012Aurasium}. In general, a user privacy concern often has its mutability. In the real world, some sensitive context in the past may not be seen as personal and private in the future. Hence, hooks do not follow such change will become useless.

Generally speaking, there is a complete method set which covers all paths to sensitive resources in the Android Framework. And we can hook every method in it to protect user privacy. Then for different smartphone owner, where to put hooks is unified although what is done in each hook is not. However, for a particular smartphone owner, such a naive solution contains too many hooks than it actually needs. This is because hooks are placed based on the whole human being's privacy concerns rather than that particular owner's. Too many hooks could introduce large system overhead and heavily slow down apps. Hence, another important motivation is that we want to optimize the naive hook placement. And we intend to systematically put hooks and automate the manual part of hook placement in
the Android Framework for each owner.

\begin{figure*}
\centering
\includegraphics[width=0.9\textwidth,height=3.5cm]{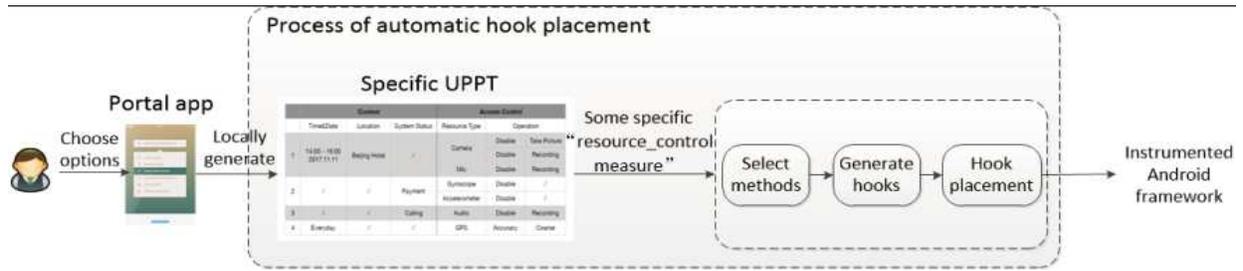}
\caption{Approach overview}
\label{figure:overviewhook}
\vspace*{-4.5mm}
\end{figure*}

\section{Definitions}\label{concept}

Since context-aware privacy protection is not only related to apps and users, but also related to the Android Framework code and hardware sensors, the proposed solution needs to see a bigger picture. To help present the bigger picture clearly, we give all the key definitions involved in the proposed solution.

\emph{Sensor Resource:} From a user's perspective, a sensor resource is one kind of sensor that may infer the user's surrounding physical environment or user activities on a smartphone. For instance, the camera is capable of inferring the user's surrounding physical environment by taking a picture or recording a video.

\emph{Sensor Data Storing Variable:} From an Android Framework developer's perspective, a sensor data storing (SDS) variable is a variable to store a piece of sensor data or a reference to the data. For instance, \texttt{notifyLocation} is one kind of SDS variable in the \texttt{LocationManagerService}, which is a main part of the Android Framework, to store GPS data.

\emph{Context:} We define a context by three aspects: the time window aspect, the location aspect, and the system status aspect. Here, the system status is meant to capture such information as (a) the foreground app name; (b) category name of the foreground app; and (c) the back stack which holds all the alive activities. For instance, a top secret meeting context may be represented by the information of the meeting start and end times and the meeting hotel name.

\emph{Privacy Concern:} A privacy concern is defined by a context, a particular set of sensor resources and a particular context-aware policy. For instance, ``in the private meeting context, disable GPS, camera, microphone'' is a privacy concern.

\emph{Policy:} A policy is a particular set of rules on how to deal with a set of sensor resources in a particular context. For instance, ``disable camera in context X'' is a simple rule that requests system to forbid one sensor resource (i.e., camera).

\emph{Sensor Data Access Method:} A sensor data access method (SDAM) is a function in the Android Framework that can read data from or write data to a SDS variable. For instance, \texttt{getLastLocation} in \texttt{LocationManagerService} will provide a GPS data to a user app.

\emph{Sensor Control Method:} A sensor control method (SCM) is a function in the Android Framework that can send a control command or call an operation interface. For instance, \texttt{scan} in \texttt{WifiNative} starts to scan available wifi network by sending ``scan'' command (to the NIC card). \texttt{takePicture} in \texttt{CameraClient} lets the camera perform the ``take picture'' operation by calling \texttt{takePicture} in \texttt{CameraHardwareInterface}.

\section{Enforcing Context-aware Privacy Policy}

\subsection{Approach Overview}
Figure \ref{figure:overviewhook} shows the overview of our approach. We develop a portal app to obtain the information about a user's privacy concerns in the User Privacy Preference Table (UPPT) through the three standard steps: (1) The portal app will show a list of sensor resources to a user, and let him choose what sensor resources need to be protected. In step 2 and 3, for the resources not selected by a user, options related to them will not be shown again. (2) For each selected resource, the app shows a list of control measures that a user intends to enforce. (3) For each control measure, the app will request a user to choose in which context the control should be enforced. Our portal app generates a specific UPPT for a specific user.

Given a specific UPPT, {\sf Prihook} will pick out the SDAMs and SCMs that are specific to the UPPT from the Potential Method Set which intends to be the union of all the individual-privacy-concern-specific (sensitive) method sets. (In Section \ref{ml}, we will describe how this set is obtained.) It automatically generates hooks and instruments the selected SDAMs and SCMs in the Android Framework.

Finally, the mobile phone manufacturer can update a instrumented Framework image to the owner's smartphone through online system update \cite{HUAWEI,SAMSUNG}.

\subsection{User Privacy Preference Table}\label{design_ppt}
Generally, there may be thousands of different kinds of user privacy concerns in the world. In this paper, we are aimed to address context-aware privacy concerns rather than arbitrary concerns. Our UPPT can represent all policies in previous works on context-aware user privacy in the Android Framework\cite{chakraborty2014ipshield,PMP, Petracca2015AuDroid,xu2015semadroid,brasser2016regulating,XposedFramework,appops}. When designing UPPT, we try to make a trade-off between user friendliness and expressiveness by two principles:

\textbf{Good coverage for users' privacy requirements.}
Although it is hard to design a table that can satisfy all people's needs, recent researches \cite{liu2014reconciling,lin2014modeling} show that a small set of privacy profiles can simplify most people's privacy decisions, which provides a promise to design such a table. A privacy profile works as a policy table which contains typical user-predefined rules on how an app can access a resource. The schema of UPPT is determined by a survey of many users' privacy decisions on their devices. We analyse all previous works' profiles \cite{liu2014reconciling,lin2014modeling,felt2012ve,lane2010survey,Krumm2009A,Consolvo2005Location,Mano2010Anonymizing,G2005Survey}, summarize context-aware concerns on sensors, and design UPPT to cover them. Although we cannot guarantee that all people are satisfied with this particular set of privacy concerns, it can work well if one's concerns are involved in our summary of previous works' profiles.

\begin{figure}
\centering
\includegraphics[width=0.4\textwidth,height=3.5cm]{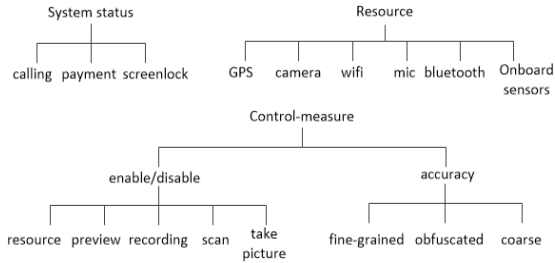}
\caption{All possible options currently implemented}
\label{figure:uppt-design}
\vspace*{-3mm}
\end{figure}

\begin{figure}
\centering
\includegraphics[width=0.45\textwidth,height=2.8cm]{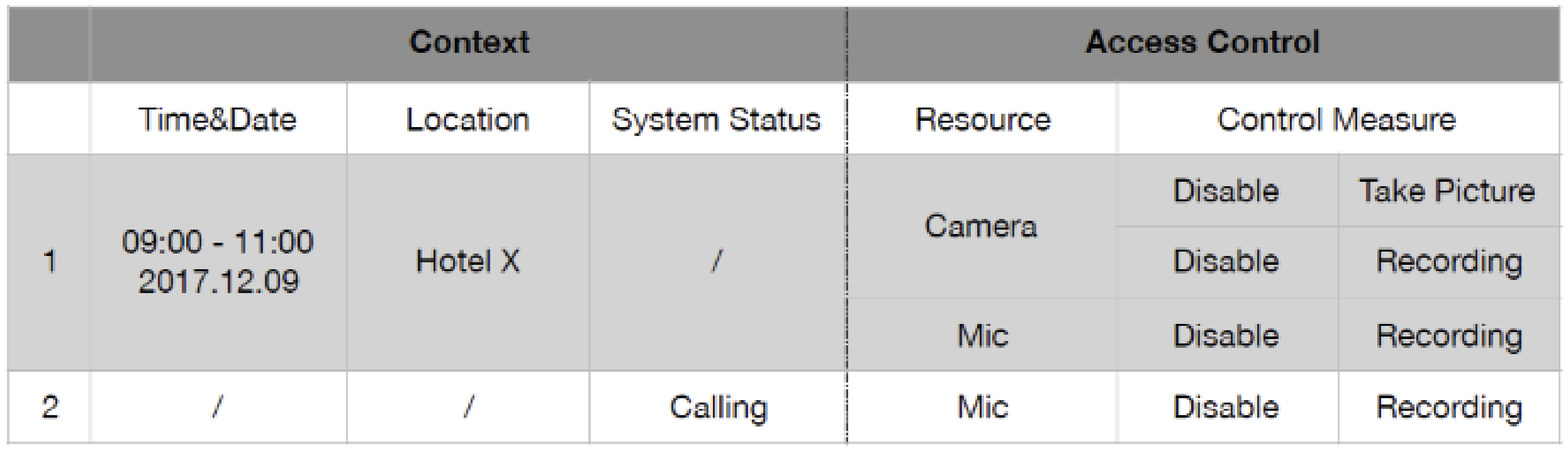}
\caption{A UPPT capturing a user's privacy concerns}
\label{figure:ppt}
\vspace*{-6.5mm}
\end{figure}

\textbf{Easy to use.}
We do not assume that a user has any background knowledge about mobile security. Hence, the items a user need to choose from the portal app should not be lengthy and cumbersome. Specifically, the UPPT schema only consists of five kinds of information, that is, system status, time, location, resource and control measure. Figure \ref{figure:uppt-design} shows all possible options currently implemented in {\sf Prihook}. Compared to existing UI of Android permissions (showing a list of technical permissions) during the app installation, our portal UI is simpler and consists of less technical terms. Figure \ref{figure:ppt} shows an example UPPT after a user has chosen items in the portal UI through the 3 standard steps. Words (designed by us) that can be chosen by a user are limited and all of them construct the UPPT lexicon. Although our UPPT policy is designed by some previous works' profiles, it can be
extended to support more policy
such as location-aware policy \cite{Kaasinen2003User,Anthony2007Privacy,Beresford2004Mix,Narayanan2001Realms,Froehlich2007MyExperience,Patrikakis2009Personalized,Ridhawi2010Policy,Wood2012Preserving,Myles2003Preserving,Freytag2009Privacy,Damiani2010Privacy,Stenneth2011Privacy,Jos2010Jano}. For instance, we
can extend our ``accuracy'' to support not only
obfuscating but also subsampling and mixing \cite{Brush2010Exploring}.

\subsection{Constructing the Mapping}\label{mapping}

In order to select the SDAMs and SCMs that are specific to a particular UPPT from the Potential Method Set (PMS), we need to build a mapping from methods in PMS to words in the UPPT lexicon.

\textbf{Challenges.} One main challenge in building the mapping from the rows in a UPPT to the PMS is that there is a large semantic gap between them. The information (e.g., policies) in UPPT cannot directly help us select methods from PMS for hook placement because policy lexicon usually does not contain any method names in the Android Framework. The policy lexicon must be user-readable so that users can fill out the UPPT. However, many if not most method names in PMS are not really user-readable. Hence, policy lexicon will not contain method names unless we stop letting users fill out UPPT. Meanwhile, it is unrealistic to assume there is a security analyst or engineer who is able to directly map the UPPT to methods in Layer 3. Such a direct mapping requires too much detailed knowledge about methods in Layer 3 and the program logic in these methods. For instance, an engineer knows a user app can get GPS data through \texttt{ILocationManager} in \texttt{LocationManagerService} and thus maps method \texttt{getlastlocation}. However, he may lack detailed knowledge about \texttt{IBlueTooth} in \texttt{Bluetooth} and thus cannot map method \texttt{startDiscovery}.

\begin{figure}
\centering
\includegraphics[width=0.5\textwidth,height=4cm]{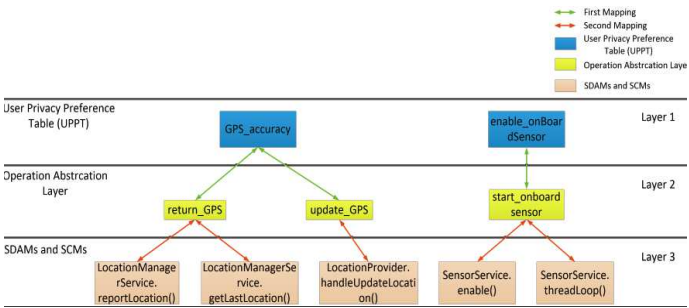}
\caption{The overview of our mapping}
\label{figure:wt}
\vspace*{-5.5mm}
\end{figure}

\textbf{Operation abstraction layer.} As shown in Figure \ref{figure:wt}, to address this challenge, our key idea is to break the mapping (from Lay 1 to Layer 3) into two mappings by introducing the Operation Abstract Layer (OAL) as Layer 2. The first mapping is from ``resource\_control-measure'' words in the UPPT lexicon (Layer 1) to abstract operations in OAL (Layer 2). The second mapping is from abstract operation in OAL (Layer 2) to methods in PMS (Layer 3). These two mappings have different properties. The first one is mostly to deal with user privacy policy and to figure out what is a sensitive operation. The second one is only focused on analyzing the source code of the Android Framework. And we can bring in automation here. The OAL brings two advantages: First, compared to the direct mapping, building a mapping from Layer 1 to Layer 2 (the first mapping) does not require detailed knowledge about methods in Layer 3. Second, When an expert (or engineer) makes a mapping, the initial large semantic gap is replaced by two smaller ones. In Section \ref{eval}, our evaluation shows that the OAL reduces the chance of making mistakes. We recruit about 100 engineers to implement the first mapping. This mapping is important because if we intend to implement a control measure on a resource, our first thought is often to look for the sensitive operations that may result in undesired access to such resource. We can place hooks in related methods to prevent such access. With the first mapping, we are able to \emph{conceptually} identify such sensitive operations.

Operation Abstract Layer (OAL) brings a new notion \emph{abstract operation}. For methods in PMS, we observe that sensitive operations can be divided into three catogories: (a) reading/write a SDS variable, which changes sensor data value in a system service; (b) receiving/sending a SDS variable through Inter-Process Communication (IPC), which may leak sensor data to a user app; (c) directly calling interfaces provided by hardware module or indirectly calling them by sending commands, which changes the status of a sensor hardware or leads to a hardware action (e.g., camera focusing). We find that we are able to identify the abstract, high-level semantics of these sensitive operations. We call these high-level semantics/words \emph{abstract operation}. For example, ``return\_GPS'' is such an abstract operation which means sending GPS data to a user app. It corresponds to two sensitive methods (i.e., \texttt{getlastlocation},\texttt{handleLocationChanged}). All abstract operations construct the OAL. Identifying the abstract operations is currently a manual procedure.
We use our knowledge about the Android Framework (i.e., the knowledge about the work flow of each system service and how each sensor works) to identify about 70 abstract operations on sensor resources. For example, there are 4 abstract operations on GPS. In order to prevent undesired access, sensitive methods which perform these operations may be hooked.

\begin{figure}
\centering
\includegraphics[width=0.47\textwidth,height=5cm]{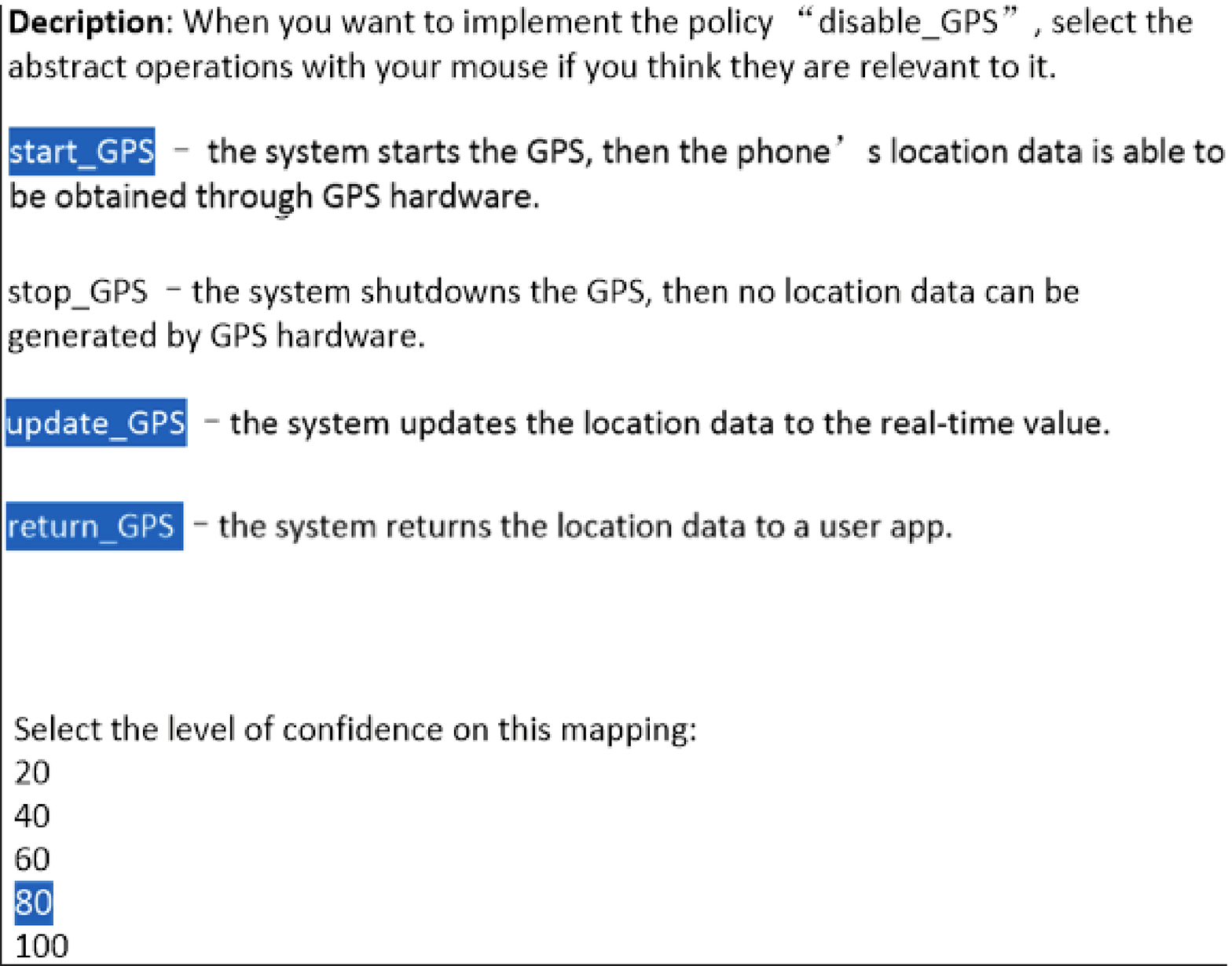}
\caption{A mapping task for GPS}
\label{figure:fmtask}
\vspace*{-7.5mm}
\end{figure}

\textbf{Layer 1 to layer 2 mapping.} We build the first mapping with three steps: (a) We write an explanatory note for each abstract operation, and recruit 100 Android engineers to take 2 days to finish a series of mapping tasks. (b) For each mapping task, every engineer separately identifies what abstract operations should be involved when he intends to implement specific control measures on a specific sensor resource and marks a confidence level (the score ranges from 20 to 100). (c) We do not take identification result into acount unless it is made with a confidence score of 80 or higher. We discuss the result and build a mapping only when more than half number of identifications are all agreed that operation should be involved. The engineers we recruited have at least 2 years experience in Android security development (e.g., Android app security or SEAndroid), which makes them qualified to implement the first mapping. Figure \ref{figure:fmtask} shows an example task an app developer identifies operations such as ``start\_GPS'' and ``return\_GPS'' when he wants to disable GPS.

\textbf{Layer 2 to layer 3 mapping.}\label{static_analysis} We build the Layer 2 to Layer 3 mapping through a unique human-in-loop iterative process. The objective is to bring two desired results closer to discovery with each iteration. The two desired results are (1) a ``bucket'' of methods (SDAMs or SCMs) for each abstract operation, and (2) a set of keywords for each abstract operation.

Before the iterative process starts, we initialize the two desired results as follows: we assign an empty bucket to each abstract operation; we recruit Android engineers to identify a tentative set of keywords for each abstract operation. We leverage the locality property to make this task manageable for the recruited engineers. In particular, given an abstract operation, the locality property ensures that an engineer only needs to be concerned with a particular system service and a \emph{subset} of methods in PMS.

After the iterative process starts, the workflow of each iteration is as follows: (Step 1) it runs Algorithm 1 to statically analyze the source code of each method in PMS for the purpose of checking whether a keyword in desired result \#2 appears in any methods; (Step 2) it uses the findings obtained in Step 1 to regenerate a bucket of methods for each abstract operation through keyword lookup; (Step 3) it asks the recruited engineers to examine the methods inside each bucket, spot misplaced methods, and used the misplaced method as a ``clue'' to refine desired result \#2.

At the end of each iteration, both of the two desired results usually get closer to discovery (of ground truth). The whole iterative process will end when a new iteration does not result in any changes to either of the two desired results.

As mentioned, the second mapping is only to analyze the source code of the Android Framework, so we intend to semi-automate the procedure through some keywords rather than pure manual efforts.
Algorithm \ref{algo} shows our static analysis. It takes three inputs: modules containing source code of SDAMs and SCMs, a set of names of SDAMs and SCMs, an abstract operation list (AO list). And it extracts sub call-graph from the whole call-graph of the Android Framework in a bottom-up way. It analyzes a method after all its callees have been analyzed. Overall, Algorithm \ref{algo} firstly identifies relevant \emph{keywords} from each method in PMS. Then, it uses keyword lookup (against desired result \#2) to obtain a subset of abstract operations. Such a subset indicates the abstract operations which the method being analyzed may perform. The \texttt{GATHER\_KEYWORDS} in Algorithm \ref{algo} is used to gather keywords in a SDAM (or SCM). It scans the source code of a method and searches for keywords in it.

\begin{algorithm}

  \caption{Static Analysis on Methods }
  \label{algo}
  \scriptsize
  \KwIn{(1):$M$:Modules contain source code of SDAMs and SCMs,
  (2):$N$:A set of names of SDAMs and SCMs,(3):$Ao$: An abstract operation list}
  \KwOut{For each n $in$ N, a set of abstract operations in OAL}
  Construct the call graph G of the module M

  L:= the list of vertices of G

  \ForEach{$f\in L$}
  {
    S := GATHER\_KEYWORDS(f);

    \ForEach {g such that f calls g}
    {
        \uIf{ $g\in PMS$}
        {
             S := S $\cup$ Keywords(g)
        }
    }

    Keywords(f) := S

    SOPS := LOOKUP\_ABSTRACT\_OPERATION(S, Ao)

    Result(f) := SOPS
  }
  return Result(f);

\end{algorithm}

\begin{figure}
\centering
\includegraphics[width=0.48\textwidth,height=2cm]{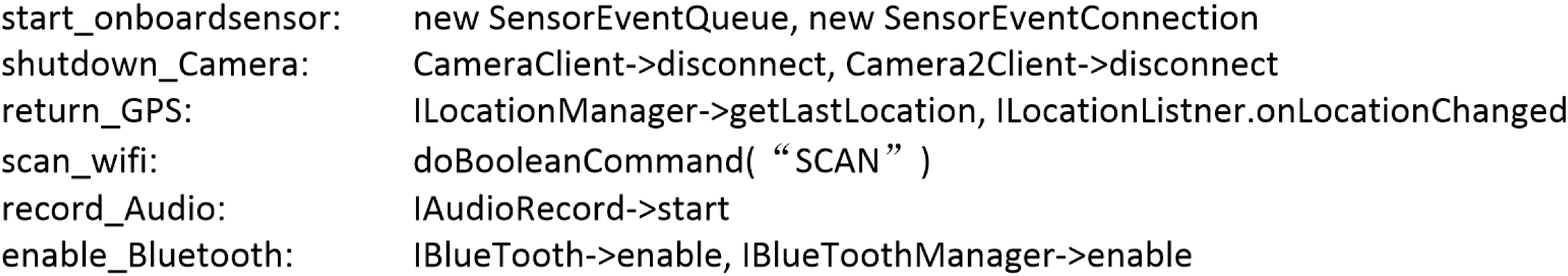}
\caption{A portion of keywords for abstract operations}
\label{figure:rules}
\vspace*{-6mm}
\end{figure}

Figure \ref{figure:rules} shows a portion of desired result \#2 used in Algorithm \ref{algo}, with abstract operations on the left side and keywords on the right. These keywords are extracted from semantic information in source code of methods in PMS. Unlike semantics in Linux kernel, most of which are manipulations of kernel data structure (e.g., inode and task\_struct), semantics in the Android Framework are more diverse and comprehensive. Based on our observations of the iterative process, we find we can use 4 kinds of semantic information as our keywords: (1) data type of a SDS variable; (2) an IPC interface between user apps and a system service; (3) an interface provided by a hardware module; (4) a macro definition or constant string which indicates a command for hardware modules. By our human-in-loop iterative process, it takes us two weeks to obtain about 200 keywords in desired result \#2.

\subsection{Selecting Methods from the PMS}
Given a user-provided UPPT, {\sf Prihook} firstly extracts ``resource\_control-measure'' words in it. Then, it directly selects a set of specific methods \#TM through searching the mapping in Figure \ref{figure:wt}. {\sf Prihook} leverages the set \#TM to extract a sub call-graph of the Android Framework. Then, for each abstract operation in each call chain, {\sf Prihook} checks whether there exists two or more methods that perform a same abstract operation. If it does, {\sf Prihook} only picks the deepest method and removes others for that abstract operation. After the ``pick-and-remove'', each abstract operation in each call chain has only one corresponding method. All construct the final hooked method set \#FM.

\subsection{Generating Hooks for Selected Methods}

\vspace*{-3mm}
\begin{lstlisting}[
caption={Hook template},
label={list:hook_template},
]
//check policy
String allowed_level = mContextAwarePolicyService.checkPolicy(notifyLocation, Binder.getCallingPid(), Binder.getCallingUid());
if (allowed_level == "DISALLOW"){
	return null;
}else if (allowed_level == "OBFUSCATE"){
  //control data accuracy
	notifyLocation = mContextAwarePolicyService.obfuscate(notifyLocation);
}
\end{lstlisting}

{\sf Prihook} uses a hook template to generate hooks for selected methods. The List \ref{list:hook_template} shows an example of a generated hook in \texttt{handleLocationChangedLocked} in \texttt{LocationManagerService}. It is a typical if-else code style. The hook generated by {\sf Prihook} is able to check policy for a resource (Line 2) or control the accuracy of a resource that a user app can obtain (Line 7).

\subsection{Placing Hooks}
Placing hooks is straightforward. For each selected method, {\sf Prihook} places hooks at the first line in it.

\subsection{Discovering a PMS in the Android Framework}\label{ml}
\renewcommand{\thefootnote}{\fnsymbol{footnote}}
\setcounter{footnote}{-1}

Our goal is to discover a method set from which {\sf Prihook} can select a particular subset to put hooks. In order to avoid ``the bypass'' mistake, it intends to be the union of all the individual-privacy-concern-specific (sensitive) method sets.

One traditional way is a manual inspection of the Android Framework. With a detailed understanding of the Android Framework code base, we can obtain a set of hand-picked SDAMs and SCMs. However, considering the large body of the Android Framework (e.g., 384,296 methods in Android 6.0), the hand-picked approach is impractical and error-prone.

Another possible way is to use the dynamic taint tracking technique. However, we cannot use this technique due to two reasons. First, a SCM cannot be discovered by dynamic taint tracking techniques. A SCM is a method which can send a control command. It does not contain any SDS variable. Hence, we cannot discover these methods by tainting SDS variables. Second, user privacy highly rely on Android native services, which are implemented in C++. For instance, a user app can obtain onboard sensor data and pictures directly from native services (e.g., \texttt{SensorService}, \texttt{CameraService}) and infer user privacy. However, dynamic taint analysis tools/techniques like TaintDroid \cite{Enck2015TaintDroid}, TaintART \cite{sun2016taintart} and NDroid \cite{qian2014tracking} can only help us monitor how sensor data flow through Android Java services and JNI runtime libs during execution. A native service runs as a Linux process and can be executed directly without the interpretation of Android Java virtual machine (i.e., Dalvik or ART). Hence, tools such as TaintDroid fail to analyze native services. While TEMU \cite{Song2008BitBlaze} is able to do dynamic taint analysis on binary code, it requires a laborious and tedious manual inspection of the Android Framework to find all possible \texttt{source} and \texttt{sink}. How to automate such finding process is non-trival.

Further, our problem of finding SDAMs is essentially \emph{data dependent} and needs a backward taint analysis. The existing static analysis techniques are not feasible to solve this problem due to the following reasons:

First, the data dependence requires us to focus on methods that read/write a SDS variable. A static analysis using control flow graph (CFG) does not contain such information.

Second, compared to dynamic taint analysis, the accuracy of static taint analysis is much lower. It suffers a huge number of false positives especially in the case of static backward taint analysis, which only knows \texttt{sink} and needs to find all \texttt{source}. To achieve a high accuracy analysis, we need a precise modeling
of the runtime execution for the Android Framework (consisting of Java and C++ code). However, how to build such a runtime model for precise static taint-analysis is still an open problem for the community. Although machine learning used by {\sf Prihook} does not guarantee to identify every SDAM and SCM, it is a practical best-effort solution for this problem. In cross-validation,
our machine learning achieves a precision of over 96\%, which means that the PMS generated by our approach greatly
reduces the risk of missing
sensitive methods for hook placement.

Third, besides the risk of being imprecise, the existing tools such as FlowDroid \cite{arzt2014flowdroid} and IFDS \cite{sagiv1996precise,naeem2010practical}
have been beset with several problems. When leveraging FlowDroid to do the static taint-analysis, we meet a problem of missing \texttt{source}. For instance, updating the current GPS location to a system service does not need a meaningful \texttt{source} as defined by FlowDroid.
\cite{hu2016third} has shown that even in the case of marking all methods in a lib as \texttt{source}, the forward taint analysis for that specific lib is not scalable. We believe it also cannot be scalable for the Android Framework if we mark all methods in it as \texttt{source}. IFDS is not designed for the Android Framework and is not trivial to be ported.

Hence, we decide to leverage machine learning to discover SDAMs and SCMs in both Java and C++. It is an automated and scalable approach  and keeps a good correctness (e.g., about an average precision of 96\%). Specifically, we leverage \emph{support vector machine} (SVM) to discover such a method set by classifying methods into two categories: SDAM\&SCM and the others.

\textbf{Our approach.}  We group Framework methods into 2 parts: the first group is a training data set which we manually annotate from Framework code, the second group is a test set that we do not know whether a method is a SDAM (or SCM) or not. We use \emph{supervised} learning to train a classifier on a
relatively small subset of manually-annotated training examples. The training set is much smaller than the test set (i.e., about 0.8\% in our case). We build a feature database for our classification (see \textbf{Feature selection} shortly). 3 students manually annotated 1000 methods from the Android Framework as a training set and abstract features from them to identify SDAMs and SCMs. We use \emph{Sequential Minimal Optimization} (SMO) implementation in RStadio with a radial kernel to realize our classifier.

\textbf{Feature selection.} We use 143 semantic features to classify methods in the Android Framework. Our evaluation shows that a combination of all features can provide enough information for classifying. We group them into 4 major categories:

\emph{Class Name} A method in a specific class contains a specific string like ``location'' and ``machine''. Many service classes handling sensor data contain an explicit name of a sensor (e.g., \texttt{LocationManagerService}).

\emph{Method Name} The name of a method contains or starts with a specific string like ``camera'', ``callback'' and ``start''. Many SDAMs directly contain an explicit name of a sensor, such as \texttt{getCameraInfo}, which can be used to infer an access method.  Callback functions (e.g., \texttt{callbackFaceDetection}) are widely applied to handle sensor data, which can be used to infer a method.

\emph{Parameter Name} Parameters of a method contain a specific string like ``buffer'', ``event''. Many parameters are organized in a specific data structure (e.g., \texttt{sensors\_event\_t}).

\emph{Return Value Type} Return value type of a method contains a specific string like ``void''. SDAMs and SCMs normally do not return a void value, which helps us exclude parts of methods.

The reason for why these features work is that the source code in the Android framework often follows a
certain regular coding style or contains duplicated parts of one method's
implementation when implementing another, which leads to a certain degree of
regularity and redundancy. A machine learning approach like ours can discover and learn it.

\begin{table*}[th]

\centering
\fontsize{6.5}{8}\selectfont
\caption{The number of SDAMs and SCMs discovered in the Android Framework}
\label{mldiscover}
\resizebox{\textwidth}{9mm}{
\begin{tabular}{|c|c|c|c|c|c|}
\hline
Android version    & Framework Methods & Package Methods & Total Methods & Total SDAMs and SCMs & SDAMs and SCMs in system services and apps \\ \hline
Android-4.2.2\_r1  & 82627             & 39912           & 122539        & 1260                 & 1146                                       \\ \hline
Android-5.1.1\_r14 & 285413            & 71980           & 357393        & 2491                 & 2156                                       \\ \hline
Android-6.0.1\_r1  & 302610            & 81686           & 384296        & 2927                 & 2670                                       \\ \hline
\end{tabular}}
\vspace*{-3.5mm}
\end{table*}

\section{Evaluation}\label{eval}

\subsection{Constructing Six Representative UPPTs}\label{6uppt}
{\sf Prihook} leverages user privacy concerns to optimize hook placement. In order to do a series of evaluation, we should firstly construct a set of typical UPPTs as our evaluation foundation. We recruit six volunteers whose professions are different and let them fill their UPPTs.  Considering the page limit, the appendix only shows two of them. {\sf Prihook} separately instruments six Android Framework based on six UPPTs. And we add a system service named \texttt{ContextAwarePolicyService} in each new code base and flush six different Android images separately for each volunteer's phone. The system service in each image can check policy and do data obfuscation.
\begin{figure}
\centering
\includegraphics[width=0.45\textwidth,height=3.6cm]{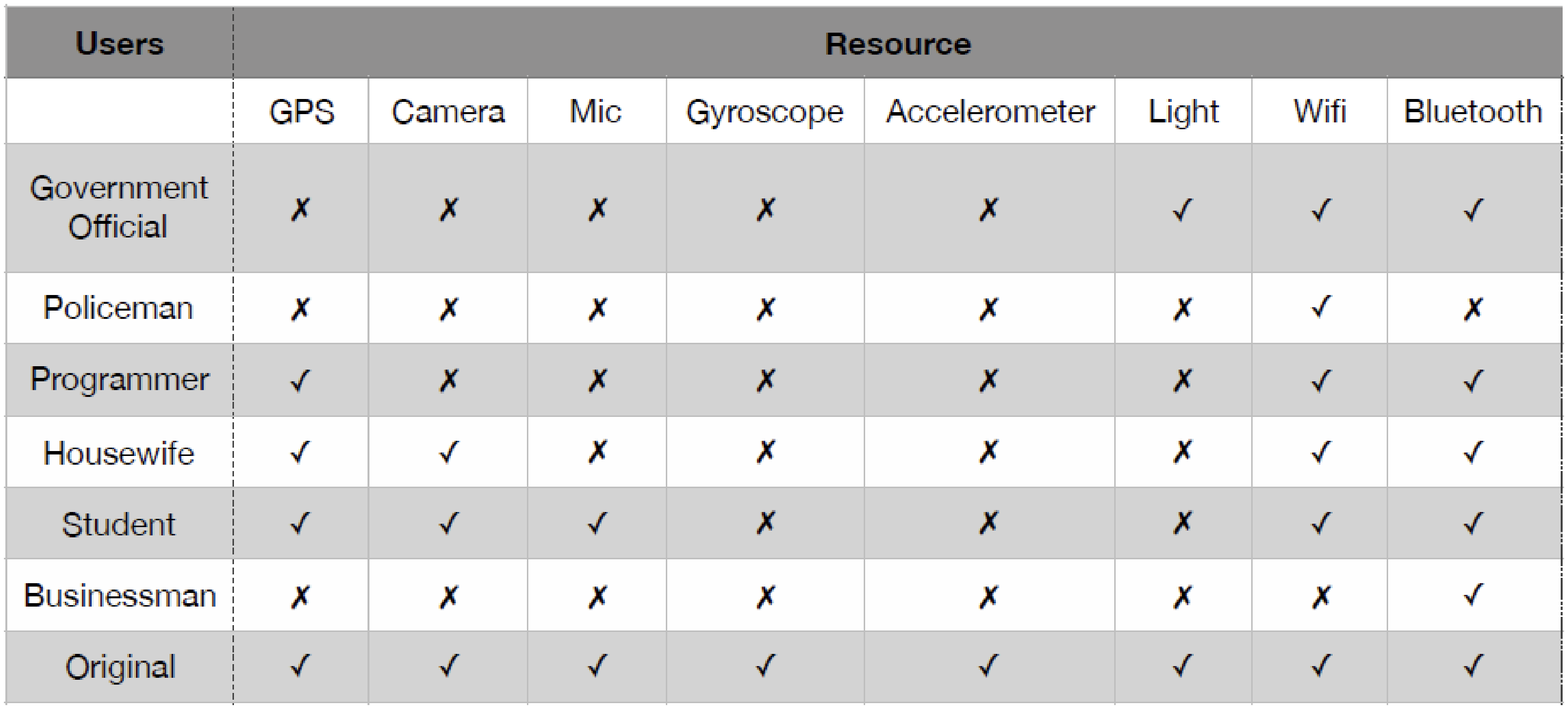}
\caption{The result of attacks on sensitive resources}
\label{figure:attack}
\vspace*{-5.5mm}
\end{figure}

\subsection{Privacy Protection}

\begin{figure}[H]
\centering
\includegraphics[width=0.45\textwidth,height=3cm]{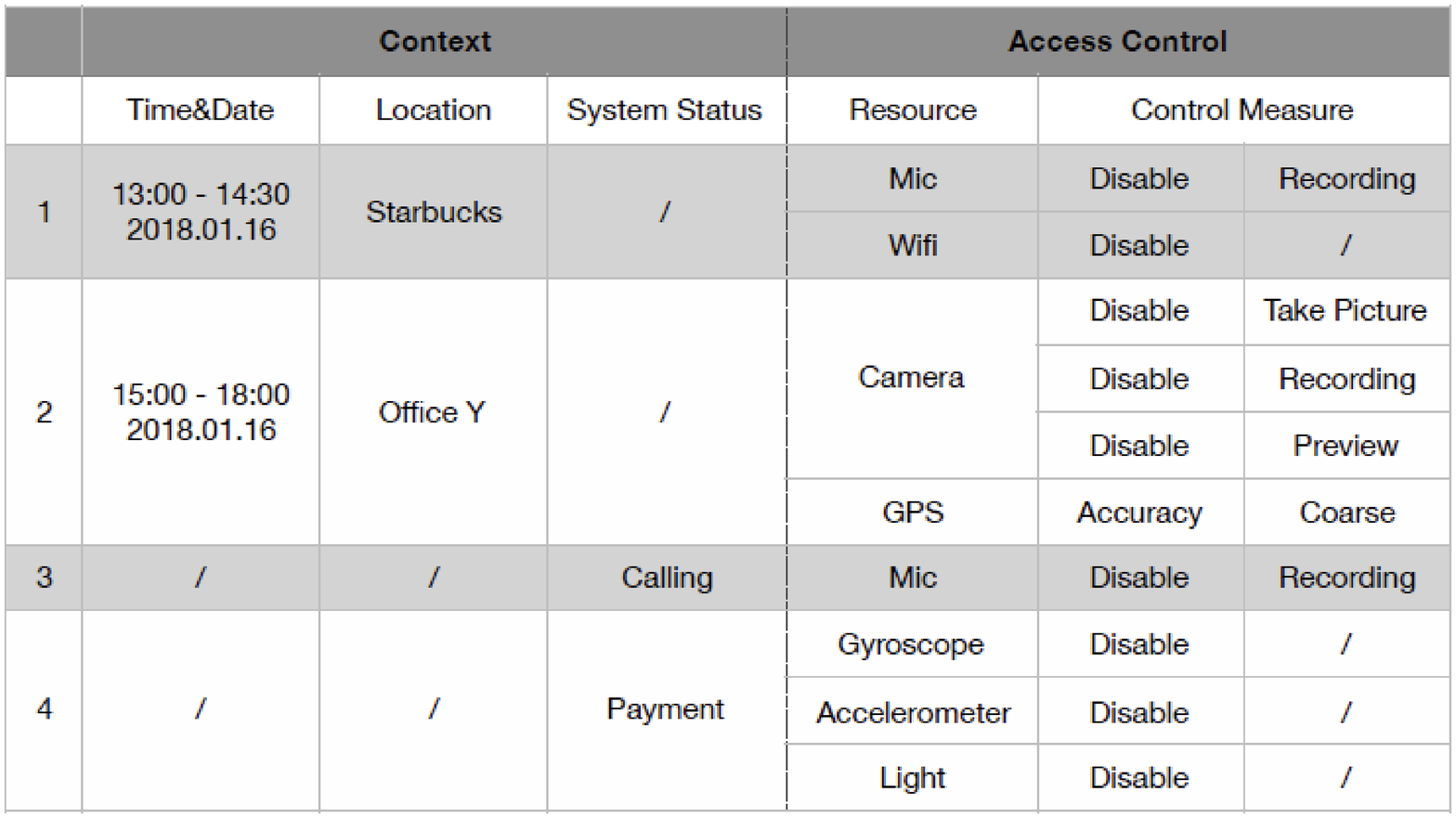}
\caption{The location\&wifi-sensitive UPPT filled by a businessman}
\label{figure:uppt6}
\end{figure}

To evaluate the defense effectiveness of {\sf Prihook} against privacy leaks, we construct a malicious app which is authorized with all permissions. It is capable to obtain sensitive resources by (a) calling SDK APIs with a complete permissions list; (b) using Java reflection to access hidden API methods; (c) directly communicating with services. We summarize all sensitive resources from 6 UPPTs. For each resource, we let 6 volunteers separately use their phones under specific context defined by their UPPTs. Then, our malicious app leverages the three ways separately to obtain sensitive resources under the specific context. As shown in Figure \ref{figure:attack}, our result shows that {\sf Prihook} can prevent attackers from obtaining a sensor resource if it is not allowed in the UPPT. For instance, the malicious app cannot obtain the audio data as there is a privacy concern about microphone in Figure \ref{figure:uppt6}.

\subsection{Impacts on User Apps}

\begin{table*}[th]
  \centering
  \fontsize{4.8}{6}\selectfont
  \caption{Elapsed time for the user app to get sensor data on the payment-sensitive UPPT without/with Prihook.}
  \label{tab:TimeCost}
  \resizebox{\textwidth}{14mm}{
    \begin{tabular}{|c|c|c|c|c|c|c|c|c|c|c|c|c|}
    \hline
    \multirow{3}{*}{User apps}&\multicolumn{2}{c|}{GPS}&\multicolumn{2}{c|}{Camera}&\multicolumn{2}{c|}{Onboard sensors}&\multicolumn{2}{c|}{Wifi}&\multicolumn{2}{c|}{Audio (mic)}&\multicolumn{2}{c|}{Bluetooth}\cr\cline{2-13}
    &Without Prihook&With Prihook&Without Prihook&With Prihook&Without Prihook&With Prihook&Without Prihook&With Prihook&Without Prihook&With Prihook&Without Prihook&With Prihook\cr
    &(ms)&(ms)&(ms)&(ms)&(ms)&(ms)&(ms)&(ms)&(ms)&(ms)&(ms)&(ms)\cr
    \hline
    Financial App A	  &3143	&3156	&6028	&6042	&2565	&forbidden	&5458	&5472   &not used	&not used	&not used	&not used\cr\hline
    Financial App B	  &2990	&3002	&6160	&6173	&2513	&forbidden	&5432	&5464   &not used	&not used	&not used	&not used\cr\hline
    Texi App 	  &4202	&4218	&not used	&not used	&1890	&forbidden	&5130	&5155   &6552	&6565	&not used	&not used\cr\hline
    Video App 	      &3075	&3088	&8441	&8465	&not used	&forbidden	&5601	&5614   &not used	&not used	&not used	&not used\cr\hline
    Social App A  &3098	&3112	&7348	&7361	&1918	&forbidden	&5221	&5052   &6069	&6620	&not used	&not used\cr\hline
    Music App B  &2524	&2543	&not used	&not used	&2321	&forbidden	&5014	&5031   &not used	&not used	&9342	&9357\cr\hline
    Map App A  &2982	&3003	&not used	&not used	&2037	&forbidden	&5102	&5123   &6231	&6255	&not used	&not used\cr\hline
    Gaming App A  &3122	&3136	&not used	&not used	&2611	&forbidden	&5002	&5026   &not used	&not used	&not used	&not used\cr\hline
    \end{tabular}}
    \vspace*{-3.5mm}
\end{table*}

In order to test if our hooks cause crash or slowdown of user apps, we use the Android automated testing tool MonkeyRunner to download 150 user apps from the Android Market. For each UPPT, we use MonkeyRunner to install and run these 150 user apps. These apps includes financial apps (e.g, bank apps), taxi apps, sports apps and music apps. Almost all these apps need some sensors such as GPS and bluetooth to function correctly. We design touch events to emulate user activities on each app and test whether it will crash and how much time it costs a user app to get sensor data. The cost time refers to the duration from the time when a documented SDK API is called, to the time when a user app gets the data.

Table \ref{tab:TimeCost} shows how much time it costs a user app to get sensor data on top 8 apps with the payment-sensitive UPPT. For each sensor data allowed by UPPT, a mount of delay (about several milliseconds) may happen. This is because a SDK API (e.g., \texttt{getLastLocation}) often results in invoking several hooks. On average, the time of checking policy and data obfuscation (i.e., randomizing data in our implementation) in each hook is less than two millisecond on Nexus 5, which is negligible. There is a Denial of Service (DOS) for a map app on onboard sensor resources when our hooks execute payment-sensitive UPPT policies that disable onboard sensors. However, this DOS is caused by user policies rather than by inappropriate hooks because policies disallow such an access. It seems that the app relies on onboard sensor data to provide smartphone's moving direction. If the data access is disabled, it cannot function well. We also test a location-sensitive UPPT and find some gaming apps work improperly. This is because these apps heavily rely on the real-time data to determine user movements. We also evaluate the number of hooks for each UPPT in Table \ref{totalhooks}.

\subsection{Precision and Recall of Machine Learning}

We expect the machine learning can discover a potential method set of SDAMs and SCMs. This approach can make sense only if it keeps a very high precision of discovering methods that are related to sensor resources. One approach to confirm this is to manually identify the correct set of all SDAMs and SCMs in the Android Framework and compare it with the result list generated by {\sf Prihook}. However, the code base of the Android Framework is really large which makes this approach impractical. Hence, with our best-effort, we take the cross validation to access such effectiveness.

\begin{table}[]
\centering
\caption{Precision and recall of our machine learning}
\label{my-label}
\begin{tabular}{|c|c|c|}
\hline
Android version    & Precision & Recall \\ \hline
Android-4.2.2\_r1  & 0.9688    & 0.932  \\ \hline
Android-5.1.1\_r14 & 0.9705    & 0.945  \\ \hline
Android-6.0.1\_r1  & 0.9656    & 0.927  \\ \hline
\end{tabular}
\vspace*{-7mm}
\end{table}

We apply a standard approach of a ten-fold cross validation. Specifically, we randomly pick 5100 methods from the Android Framework, annotate and use them both as training and tests data set. The validation process works as follows: The hand-annotated mehotds will be randomly divided into ten equally-sized groups. We train the machine learning classifier on nine of them and classify the remaining one. The whole approach will be repeated ten times. We calculate the average precision and recall about our classifier. We do the cross validation on 3 different Android versions. Table \ref{my-label} shows the final results. Since the test data set is randomly picked, the precision and recall should carry over to
the entire Android Framework with high probability. Some helper functions such as \texttt{load\_audio\_interface} are misclassified as SDAMs by {\sf Prihook}. Such a method helps \texttt{AudioFlinger} load hardware module and does not contain any SDS variable. Besides, we find that {\sf Prihook} may leak some SDAMs such as \texttt{updateLinkProperties}. However, we think such a mistake will not lead to a privacy violation because {\sf Prihook} discovered the method \texttt{addressUpdated} which calls \texttt{updateLinkProperties}. In other words, they are in a same call chain. We think this mistake happened because these two method names do not provide enough code redundancy that {\sf Prihook} can use.  However, after adding some new features to our machine learning, {\sf Prihook} can find such a case.

Except for SVM, we test other classification algorithms which fall short of discovering the PMS. We find that SVM significantly outperforms Naive Bayes \cite{John2013Estimating} on discovering SDAMs. The possible reason could be the violation of the Bayes assumption about the conditional independence of the feature sets. The pruned C4.5 decision tree \cite{Quinlan1992C4} gets a precision and recall of about 70\%. We think that the main problem
with a rule set is its lack of flexibility. A rule tree would often
include a rule mapping all \emph{set} methods to SDAMs. However, some SDAMs do not start with \emph{set}. Besides, not all methods that
start with \emph{set} are actually SDAMs. With an SVM, such aspects that are usually correct, but
not always, can be expressed more appropriately by shifting
the hyper-plane used for separation. Besides, we also try to use \emph{semi-supervised} learning to train a classifier such as co-training \cite{Blum1998Combining} with xgboost \cite{XGBoost}, which does not get an accuracy as high as SVM.

\subsection{Application across Android Versions}\label{acrossv}

\begin{table}[]
\centering
\fontsize{8}{10}\selectfont
\caption{The number of hooks placed by each user's UPPT on Android 4.2.2}
\label{totalhooks}
\begin{tabular}{|c|c|c|}
\hline
\textbf{UPPT}     & \textbf{The number of hooks} \\ \hline
Location-sensitive & 47                                          \\ \hline
Location\&onboardsensor-sensitive              & 81                                          \\ \hline
Location\&payment-sensitive          & 58                                          \\ \hline
Calling-sensitive           & 23                                          \\ \hline
Payment-sensitive             & 19                                          \\ \hline
Location\&wifi-sensitive         & 92                                          \\ \hline
\end{tabular}
\vspace*{-6mm}
\end{table}

To test how well it performs for different Android versions, we use machine learning to discover the PMS for 3 Android versions. As shown in Table \ref{mldiscover}, our classifier detects the changes in different
API versions well. It reliably finds new SDAMs and SCMs which changes a little bit (e.g., \texttt{startscan} in Android 5.0 and \texttt{startDelayedScan} in Android 6.0). It is worth noting that for some completely new and previously unanticipated methods (e.g., \texttt{updateMonitoring} in \texttt{Receiver} in Android 5.0 or newer versions), {\sf Prihook} is able to discover
them.

\section{Related Work}
\textbf{Privacy Protection} There are many solutions to enhance Android with context-aware hooks \cite{chakraborty2014ipshield,PMP, Petracca2015AuDroid,xu2015semadroid,brasser2016regulating,XposedFramework,appops}. SmarPer \cite{olejnik2017smarper} and CRePE \cite{conti2010crepe} address common user privacy concerns through hook placement. They start from {\em permission-checks} to place hooks. However, as mentioned in Section \ref{intro}, for context-aware privacy protection, adding hooks in {\em permission-checks} is not a good choice. Protect My Privacy (PMP) \cite{PMP} lets users deny or fake access to their private data from any app or any 3rd party library based on Xposed Framework. PMP does not clearly claim where it starts from for hook placement. It involves 8 hooked methods which include \texttt{requestLocationUpdates}. However, adding hooks in this method can lead to ``no-isolation'' mistakes, because its instance runs in a user app process. To the best of our knowledge, {\sf Prihook} is the first work to place hooks based on specific context-aware user privacy concerns. Compared to the existing works, its hook placement is personalized.

Another related work is the research of the permission mapping, that is, a mapping between an API call and a permission that a user app must be authorized. PScout \cite{Au2012PScout} uses static reachability analysis for permission-check APIs and creates a permission mapping. This mapping can be used to do permission analysis \cite{Grace2012Unsafe}, compartmentalize third-party code \cite{Shekhar2012AdSplit} and study developer behaviors \cite{Vidas2011Curbing}. Unlike Pscout, Revisiting \cite{backes2016demystifying} uses object-sensitive pointer resolution to generate a more precise call-graph on its static runtime model. Kratos \cite{shao2016kratos} is focused on identifying inconsistency on Android permission system by forward analysis.

Besides, to detect or prevent private data leakage, many researchers use static or dynamic taint tracking techniques on the Android Framework to obtain information flow. Epicc \cite{Octeau2013Effective} creates specifications for each inter-component communication (ICC) \emph{source} and \emph{sink}. FlowDroid \cite{arzt2014flowdroid} generates per-component lifecycle models so that it can help understand when, how, and what data travels through the Android Framework. AndroidLeaks \cite{Gibler2012AndroidLeaks} and DroidSafe \cite{gordon2015information} also work for the same purpose. TaintDroid \cite{Enck2015TaintDroid} customizes dalvik virtual machine to achieve taint storage and taint propagation so that it can track information flows of sensitive data. TaintART \cite{sun2016taintart} is similar to TaintDroid, but it is mainly focused on Android Art runtime machine. NDroid \cite{qian2014tracking} uses dynamic taint tracking to analyze the Android JNI information flow. BlueSeal \cite{Shen2014Information} is an extension to the Android permission mechanism that characterizes the implicit interactions
between data and APIs protected by standard permissions.

\textbf{Automated Hook Placement}   TAHOE \cite{ganapathy2005automatic} provides a technique to automate the hook placement in Linux kernel. They manually identify a set of idioms of security sensitive operation. With this information, they use static analysis to summarize the operations that hooks protect and the operations that a Linux module performs. Then, placing hooks is to match operations in the Linux module with hooks that mediate the operations. \cite{muthukumaran2012leveraging} presents an automated hooking tool that can be used to put authorized hooks for X server and postgresql. They leverage the user input requests to statically analyze the server source code. It can automatically group accesses of structure members into operations and put hooks on them. The above two works do not start from specific context-aware user privacy concerns for hook placement. However, the hook placement in {\sf Prihook} is personalized.

\section{Conclusion}

In this paper, we propose an automated solution named
{\sf Prihook} for hook placement in the Android Framework. Based on a specific UPPT, {\sf Prihook} can select methods
from the PMS, generate and put hooks for the selected methods.

\section{Appendix}

\begin{figure}[H]
\centering
\includegraphics[width=0.45\textwidth,height=2cm]{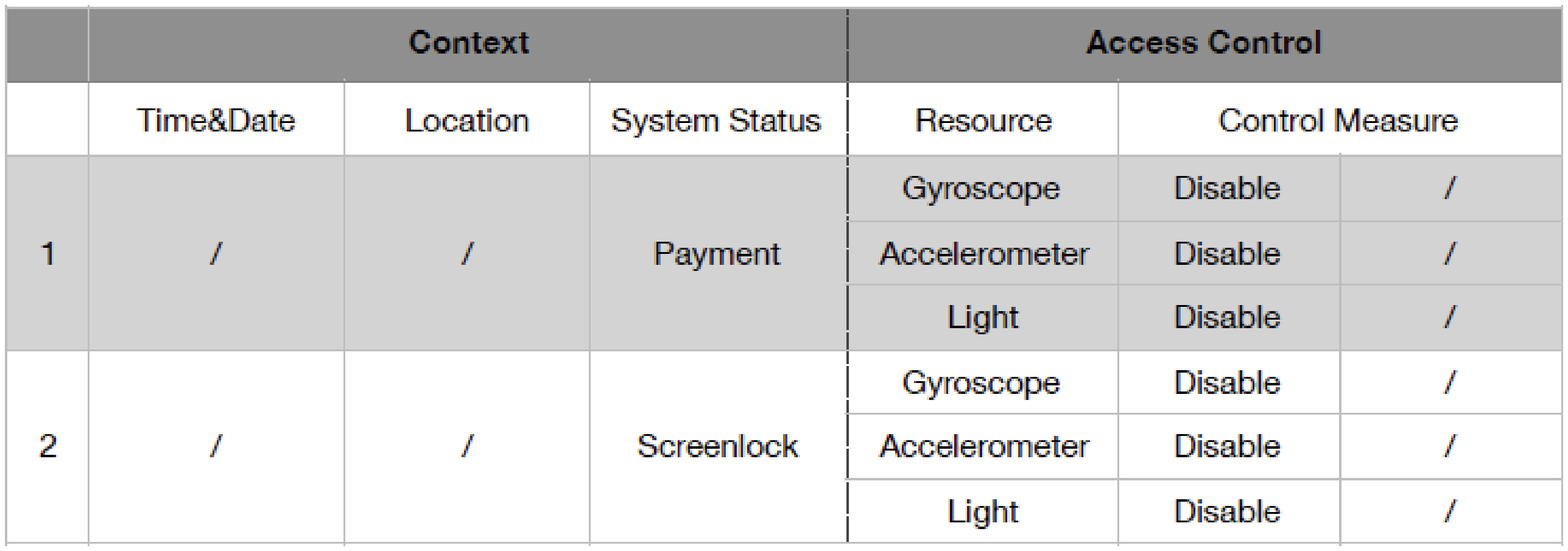}
\caption{The payment-sensitive UPPT filled by a high school student}
\label{figure:uppt5}
\vspace*{-8mm}
\end{figure}

\bibliographystyle{IEEEtran}
\bibliography{mybibfile}

\end{document}